\documentclass[onecolumn,showpacs,preprintnumbers,a4,prb,12pt]{revtex4}
\usepackage{graphicx}%
\usepackage{dcolumn}
\usepackage{amsmath}
\usepackage{amssymb}
\usepackage{bm}

\makeatletter
\def\btt#1{\texttt{\@backslashchar#1}}%
\DeclareRobustCommand\bblash{\btt{\@backslashchar}}%
\makeatother

\begin{document}


\title{Phase transition and critical properties of spin-orbital
         interacting systems}
\author{Huai-Bao Tang$^{1,2}$, Dong-Meng Chen$^{3}$, Xiang-Fei Wei$^{1,2}$,
        and Liang-Jian Zou$^{1,\dag}$ }
\affiliation{ \it $^1$~  Key Laboratory of Materials Physics, Institute of
              Solid State Physics, Chinese Academy of Sciences, P. O. Box
                1129, Hefei 230031, China }
\affiliation{\it $^2$~ Graduate School of the Chinese Academy of Sciences}
\affiliation{\it $^3$~ College of Physic Science and Technology,
                       China University of Petroleum, Dongying 257061, China}
\date{\today}

%
\begin{abstract}
Phase transition and critical properties of Ising-like spin-orbital
interacting systems in 2-dimensional triangular lattice are
investigated. We first show that the ground state of the system is a
composite spin-orbital
ferro-ordered phase. Though Landau effective field theory predicts
the second-order phase transition of the composite spin-orbital
order, however, the critical exponents obtained by the
renormalization group approach demonstrate that the spin-orbital
order-disorder transition is far from the second-order, rather, it
is more close to the first-order.
The unusual critical behavior near the transition point is
attributed to the fractionalization of
the composite order parameter.
\\

\\


$\dag$ corresponding author, Tel: 0086-551-5591428, Fax:
 0086-551-5591434, E-mail:zou@theory.issp.ac.cn
\\

\end{abstract}
\maketitle

\newpage

\section {\bf Introduction }

  In many transition-metal-oxide insulators, in addition to spin degree of
freedom, orbital degree of freedom plays important roles and leads to
complicated phase diagram and interesting phenomena, such as orbital
ordering (OO), metal-insulator transition (MIT) and colossal magnetoresistive
effect, etc. In these insulating compounds, the sign and magnitude of
the superexchange constants between magnetic ions depend on the regular
orbital occupations, i.e. the OO \cite{Kugel, Tokura, Khomskii}.
%
%
Though there are a few of debates on the microscopic origin of OO,
various experimental observations by the resonant Raman technique
\cite{Saitoh} and the resonant x-ray scattering (RXS)
\cite{Murakami} have shown the existence of long-range OO in
manganites and some other perovskite transition-metal oxides.
The importance of the orbital degree of freedom also exhibits in new
molecular compounds based on $C_{60}$ \cite{Paul},
layered fullerides, and some two-dimensional (2D) copolymers \cite{Choi}.

The low-energy physics of these orbital insulators with both spin
and orbital degrees of freedom is described by spin-orbital
superexchange interactions. The interactions are usually highly
anisotropic in orbital space, as well as in real space, due to the
Hund's rule and the anisotropy of the orbital wavefunctions. Kugel
and Khomskii \cite{Kugel} first proposed such an effective
spin-orbital model to address the complicated magnetic
structures. Such a theory incorporating the Jahn-Teller (JT) effect had
witnessed a great success in past decades in explaining the magnetic
structures of a wide range of material with the e$_{g}$ orbital
degeneracy, such as KCuF$_{3}$, LaMnO$_{3}$ and MnF$_{3}$, etc.
However recent experiments in pseudocubic perovskite
titanates\cite{Keimer}, and vanadates\cite{Ulrich} showed that in
these insulators the static OO is absent and the orbital
fluctuations and spin fluctuations should be taken into
consideration on equal foots.
Especially, the orbital degree of freedom entangled with spin degree
of freedom may lead to the violation of the Goodenough-Kanamori
rules\cite{Oles} in these spin-orbital interacting systems, and the
spin-orbital entanglement may be used to characterize different
quantum phases and their competition \cite{Chen} due to the
composite spin-orbital fluctuations, indicating that the
thermodynamic phase transition and critical properties are distinct
from those of magnetic systems with only spin degree of freedom.
The ground states and the low-energy excitations of these
spin-orbital interacting insulators have been well understood
through numerous theoretical and experimental efforts in the past
decades. While, with the increasing temperature, the spin-orbital
compounds exhibit complicated multicritical properties, how the critical
properties of the spin-orbital compounds behave near the phase
transition point is seldom investigated. Further, it is important
to understand the universal class of the spin-orbital model in
statistical mechanics and condensed matter theory.

In this Letter we present the phase transition and critical
properties of an Ising-like spin-orbital system, where the
long-range order and the order parameter of the system are composite
spin and orbital order, rather than single spin or orbital order. We
show that due to such an entanglement, the critical
exponents of the spin-orbital interacting systems are unexpectedly
different from those in magnetic systems without orbital degeneracy.
To demonstrate the general characters of the Ising-like
spin-orbital model, we first determine the spin-orbital
ordered ground state utilizing the cluster self-consistent field
(Cluster-SCF) approach, and then obtain the evolution of the
spin-orbital order parameter with temperature in the Landau
mean-field approximation. Further, the non-trivial critical
exponents, which characterize the power-law behaviors of the
order parameter of the system, are also obtained by Wilson's
renormalization group (RG) technique.
And in the final we briefly discuss our results.
%

%
%
\section{\bf Model Hamiltonian and Ground State}

The effective superexchange interaction describing the low-energy
processes in orbital-degenerate transition-metal oxides takes the
generalized Heisenberg form \cite{Kugel}
\begin{equation}
   H=-\sum_{<i,j>}[J_{S} {\bf S}_{i}\cdot {\bf S}_{j}+ f({\bf T}_{i},
      {\bf T}_{j})+g({\bf T}_{i},{\bf T}_{j}){\bf S}_{i}\cdot {\bf S}_{j}]
\end{equation}
where $J_{S}$ , $f({\bf T}_{i}, {\bf T}_{j})$ and $g({\bf T}_{i},
{\bf T}_{j})$ represent the spin, orbital and spin-orbital
superexchange constants, respectively, which originate from the
quantum mechanical intermediate virtual processes in the strong
correlation regime.
The Hamiltonian (1) describes the spin-spin, orbital-orbital and
spin-orbital couplings with the $SU(2)$ symmetry in spin space in
the system with S=1/2 and T=1/2.
In the realistic compounds, the spin $SU(2)$ symmetry of the systems is
usually broken by the magnetic anisotropy.
Moreover, the symmetry of the orbital space is broken by the
Hund's coupling and the anisotropy of the hopping integrals
$t_{\alpha\beta}$, which depends on the shapes of the 3d-orbital
wave functions and the relative orientations of two nearest-neighbor
d-orbits. Thus the effective spin-orbital interactions in realistic
compounds are highly anisotropic.

The simplest theoretical model describing the anisotropic spin-orbital
interactions in an insulator reads
\begin{equation}
   H= - \sum_{<ij>}J S_{i}^{z}T_{i}^{z} \cdot S_{j}^{z}T_{j}^{z}
      + h\sum_{i}S_{i}^{z}T_{i}^{z}
\end{equation}
where the operator $S_{i}^{z}$ labels the z-component of the electron
spin, while $T_{i}^{z}$ denote the z-component of the orbital
pseudo-spin opertaor. In what follows we let S=$\pm$1 and T=$\pm$1.
The summation in the first term is taken over the nearest-neighbor
pairs $<ij>$;  h stands for
the conjugated field of the spin-orbital order parameter.
Such an ideal model, like the Ising model for ferromagnetism,
can serve as important reference in understanding the
essential characters of the phase transition in
the spin-orbital interacting systems.
In what follows, we consider the spin-orbital system in a 2D
triangular lattice, since more and more transition-metal oxides,
such as LiVO$_{2}$\cite{Pen} and Na$_{x}$NiO$_{2}$ \cite{Li}, are
found to be triangular lattice.
%

   To get a straight insight into the physical properties of strongly
correlated spin-orbital systems in 2D triangular lattice, we utilize
the cluster self-consistent field (SCF) approach developed recently
\cite{Zou} to extract the ground state properties of the Ising-like
spin-orbital model (2) at T=0 K. The numerical results showed that
in the 3-site cluster, the groundstate energy per site is $E_{g}=-0.5625J$,
while the macroscopic groundstate degeneracy is not completely removed
at $h=0$. Such a degeneracy is removed and the ground state is
ferro-ordering in the limit of $h\rightarrow0$.
As a comparison, the ground state of the
spin Ising model is non-degenerate ferromagnetic at $h=0$.
Nevertheless, the spin-orbital correlation functions between sites in
the cluster are ferro-ordering, while the spin-spin and the
orbital-orbital correlation functions vanish, suggesting that the
spin and the orbital degrees of freedom form an entangled
triplet, and lead to the composite ferro-order, similar
to the spin-orbital ferromagnetic order found by Zhang and Shen
\cite{Shen} in the $SU(4)$ spin-orbital model in 2D square lattice.
We notice that the degeneracy of each site in the present model is
fourfold, and the ground state is degenerate "ferromagnetic", this
is comparable with
the Potts model with $p=4$, whose critical behavior had been widely
studied by Baxter {\it et al.}\cite{Baxter} and Nienhuis {\it et
al.} \cite{Nienhuis}, and the ground state forms a p-fold degenerate
ferromagnetic state, corresponding to a uniform classical state.
However, the ground state of the present mdoel is spin-orbital entangled
with $<ST>$=1.

\section{\bf Mean-Field Theory}

 With increasing temperature until to a critical value T$_{c}$,
the composite spin-orbital order is stable against the thermal
fluctuation. We adopt the Bethe
approximation \cite{Bethe} to study the temperature evolution of
the order parameter. Considering
a 7-site cluster consisting of a central site with 6 nearest neighbours.
%
%
The intra-couplings between the central site and 6 nearest
neighbours are accurately calculated, while the inter-couplings
between the nearest-neighbour sites and other lattice sites are
treated as a mean field $h^{MF}$. In the Bethe
approximation\cite{Bethe}, we find that in agreement with the
preceding results, $<S^{z}>=0$, and $<T^{z}>=0$, however, the
average of the combination of spin and orbit, $<S^{z}T^{z}>$, is
finite and given by the self-consistent equation:
\begin{equation}
     <S^{z}T^{z}> = (z-1)Ln\frac{1+e^{2\alpha+
                             2\gamma}}{e^{2\alpha}+e^{2\gamma}}
\end{equation}
in the absence of external field $h$, here
$\alpha=J<S^{z}T^{z}>/k_{B}T$ and $\gamma=J/k_{B}T$. Moreover,
Eq.(3) gives rise to a finite critical temperature at
$T_{C}=\frac{J}{k_{B}}/8\ln(z/(z-2))$ $\simeq
0.303\frac{J}{k_{B}}$.
In fact, due to spin-orbital entanglement, the Hamiltonian (2) can be
rewritten as complete Ising-like:
\begin{equation}
   H = -J \sum_{<ij>} Q_{i} Q_{j} -h\sum_{i} Q_{i}
\nonumber
\end{equation}
with the composite operator $Q_{i}=S_{i}^{z}T_{i}^{z}$. Thus one
would expect that the phase transition and critical behavior of
the present spin-orbital interacting system quite resembles to
that of the spin Ising model.
For example, one expects that the order-disorder transition of the
present system is the second order.

To further illustrate the evolution of the composite order
parameter with temperature near the critical region, we consider
the high-temperature expansion of the free energy in the {\it
Landau} mean field (MF) approximation. The order parameter,
$M=<S^{z}T^{z}>$, is a small quantity in this situation. The {\it
Landau} free energy per site becomes
\begin{eqnarray}
   F/N & =&  -2k_{B}T \ln2 + M^{2}J+ \ln(\cosh\frac{\beta
   zM}{2})\nonumber\\
    &\simeq& -2k_{B}T\ln2 +(zJ-\frac{\beta z^{2}}{8})M^{2} +
   \frac{\beta^{3}z^{4}}{192} M^{4}
\end{eqnarray}
where N is the total number of lattice sites, and z=6 is the
coordinate number. Minimizing $F$ with respect to M gives rise to
the mean-field critical behavior of the composite spin-orbital
ferro-order :  $ M \propto (T_{c}-T)^{1/2}$, near the critical
point, and the corresponding critical temperature is given by
$T_{C}=zJ/8k_{B}\simeq 0.375J/k_{B}$, a slightly large than
T$_{C}$ obtained by the Bethe approximation.

\section{\bf Renormalization Group Approach }

Obviously, the present Curie point is overestimated in comparison
with that obtained by Bethe approximation. On the other hand, the
present {\it Landau} mean-field approximation becomes unreliable,
it underestimates the thermal fluctuations near the critical
regime, and gives incorrect critical exponents near the transition
point. It is well known that the universal characters and the
scaling laws of the thermal quantities are not exactly captured by
the mean-field theory. which appeals for the renormalization group
calculations for the model (2).

In the real-space renomalization group, the triangular lattice is
divided into the 3-site cells, which interact with each other in
the similar ways as the original one as shown in Fig.1.
The transformation procedure to the spin-orbital interactions (2) is
implemented by the RG technique proposed by Niemeijer and van
Leeuwen \cite{Niemeijer}.
In the present cell structure in Fig. 1, we define the cell spin and
the cell orbit as
\begin{equation}
     S_{i}^{'} = sign(\sum_{i=1}^{x}S_{i}); ~~~~
     T_{i}^{'} = sign(\sum_{i=1}^{x}T_{i})
\end{equation}
A cell contains 16 different internal configurations which can be
labeled by the direct product of the spin sub-configurations
$\{\sigma\}$ and the orbital sub-configurations $\{\tau\}$.
By expressing the $S_{i}$ in terms of
$S_{i}^{'}$ and ${\sigma}$ and the $T_{i}$ in terms of $T_{i}^{'}$
and ${\tau}$, the transformation of Hamiltonian between the cell
system and the site system is defined as,
\begin{equation}
    exp(H^{'}(S_{i}^{'},T_{i}^{'}))=\sum_{\{\sigma\}}\sum_{\{\tau\}}
                     exp(H(S_{i}^{'}\sigma;T_{i}^{'}\tau))
\end{equation}
with $\{ \sigma \}$ and $\{ \tau \}$ running over all internal
configurations; here the reverse temperature factor $ -\beta $ is
absorbed in the coupling constants.

To approximately calculate the renormalization equation (6), we
utilize the cumulant expansion method \cite{Niemeijer}. Spliting the
Hamiltonian (6) into an internal part $H^{0}$ and a remainder part
$V$ and treating the latter as a perturbation, one has the cell
Hamiltonian in the second-order expansion of V,
\begin{eqnarray}
 && H^{'}(S_{i}^{'},T_{i}^{'})\simeq\ln<exp(V)> \nonumber\\
 && \simeq<V>_{0}+\frac{1}{2}(<V^{2}>_{0}-<V>_{0}^{2})+O(V^{3})
    \nonumber\\
\end{eqnarray}
where
\begin{equation}
    <A>_{0} = \frac{\sum_{\{\sigma\}}\sum_{\{\tau\}}A(\sigma,\tau)
              exp(H^{0}(\sigma,\tau))}
              {\sum_{\{\sigma\}}\sum_{\{\tau\}}exp(H^{0}(\sigma,\tau))}.
\nonumber
\end{equation}
In the absence of the conjugated field $h$, the evaluation to (6)
is straightforward, and leads to a set of renormalization
equations:
\begin{eqnarray}
  K^{'} &=& 2f_{1}^{2}K+8K^{2}f_{1}^{2}(1+f_{2}-2f_{1}^{2})+3f_{1}^{2}L
           +2f_{1}^{2}M \nonumber\\
  L^{'}&=&2(1+7f_{2}-8f_{1}^{2})f_{1}^{2}K^{2}+f_{1}^{2}M \\
  M^{'}&=&8(f_{2}-f_{1}^{2})f_{1}^{2}K^{2} \nonumber
\end{eqnarray}
with
\begin{eqnarray}
    f_{1}&=&\exp3K/(\exp3K +3\exp-K ) \nonumber\\
    f_{2}&=&(\exp3K-\exp-K) /(\exp3K +3\exp-K ).  \nonumber
\end{eqnarray}
The numerical results for the nontrivial fixed points of the
nonlinear equations (8) are shown in {\it Table I}. For
comparison, the results of the Ising model with
$S=\frac{1}{2}$\cite{Niemeijer} and $1$ and XY model\cite{Rogiers}
are also presented. The comparisons of the critical exponents for
various models are shown in {\it Table II} and {\it III},
respectively.
The presence of the unstable fixed points suggests that the
existence of the spin-orbital order-disorder phase transition.

Linearizing the renormalization equations (8), and diagonalizing
the matrix, $T_{\alpha\beta}={\partial K_{\alpha}}/{\partial
K_{\beta}}$, around the fixed points, we get only one relevant
eigenvalue, $\lambda_{T}$, in the second-order cumulant expansion,
as shown in {\it Table II}.
In the presence of the conjugated field h, analogous to Ref.
\cite{Niemeijer}, we have yielded the magnetic relevant eigenvalue,
\begin{equation}
    \lambda_{H} = \frac{\partial h'}{\partial h}
            = 3f_{1}+12f_{1}K(1+2f_{2}-3f^{2}_{1})
\end{equation}
in the first-order cumulant expansion, which is also listed in
{\it Table III}.
Two corresponding "thermal" and "magnetic" exponents, obtained
through $\alpha_{T(h)}={d \ln L}/{d \ln \lambda_{T(h)}}$, are also
listed in Table II and III, respectively.
Thus, according to the scaling laws, the critical exponents of
the specific heat, $\alpha$, with respect to $T-T_{C}$ and of the order
parameter, $\delta$, with respect to the conjugated field at $T=T_{C}$
are also known, and listed in {\it Table.II} and {\it III}, respectively.
Moreover, the critical temperature can be evaluated as follows
\begin{equation}
    K_{C}(J)=K_{C}^{*}-\sum_{\alpha \neq
              n.n}(\varphi_{\alpha}^{T}/\varphi_{n.n}^{T})K_{\alpha}^{*}
\end{equation}
where $\varphi_{\alpha}$ is the left eigenvectors of the matrix
$T_{\alpha\beta}$. The result is also listed in {\it Table II}.

From {\it Table II}, one finds that the critical exponent of
specific heat $\alpha$  is far from that of the Ising model with the
second-order phase transition, and the 'thermal' exponent
$\alpha_{T}$ lies between the second-order and the first order, and
$\alpha_{T}$ is more close to the first order, implying that the present
spin-orbital order-disorder phase transition of the composite order
parameter, $<S^{z}T^{z}>$, is of a weak second order in the absence
of the conjugated field.
Such a weak second-order phase transition can be clearly seen by
rewriting the singular free energy near the critical
point\cite{Niemeijer},
\begin{equation}
     f_{sing}=A\mu_{1}^{\alpha_{T(H)}}+O(\mu_{1})
\end{equation}
where $\mu_{1}$ represents the distance from the fixed points.
Thus the order of the singularity of the free energy is solely
determined by the 'thermal' ('magnetic') exponent $\alpha_{T(H)}$.
One finds that the 'thermal' exponents are close to but larger
than unity, which are different from those of the second-order
transition in the spin-$\frac{1}{2}$ and the spin-$1$ Ising
models, as seen in {\it Table II} and {\it III}.
Meanwhile, magnetic eigenvalue $\alpha_{H}$, which is close to but
smaller than unity, suggests that the phase transition induced by
the conjugated field is weak first-order.

Interestingly, we note that in {\it Table II} and {\it III} the
present critical exponents of the spin-orbital model are close to
those of the two-dimensional quantum XY model \cite{Rogiers},
suggesting that these two models belong to the same universal
class.
Since the component of order parameter of the quantum XY model is
two-dimensional, this implies that near the critical point, the
order parameter of the present spin-orbital model decouples into
two independent components, i.e., the spin and the orbital degrees
of freedom have already disentangled. This result is significantly
different from the prediction of the mean-field theory.
Moreover, the critical temperature obtained by the RG approach is
considerably reduced, in comparison with the mean-field result,
indicating that the strong fluctuations near critical points are
considered properly. We also notice that the critical temperature
is much smaller than that of the spin Ising model. Therefore, we
conclude that the thermal fluctuations are expectedly strong, due
to the disentanglement between spin and orbital degrees of
freedom.

\section{\bf Remarks and Summary}

 As we state in the preceding, the formation of the spin-orbital
composite order parameter arises from the entanglement of the spin
and the orbital degrees of freedom at the same site.
The entanglement between the spin and orbit disappears when the
system is close enough to the critical point, and strong thermal
fluctuations entirely destruct the composite spin-orbital order.
The considerable discrepancy of the critical exponents between
the RG approach and the {\it Landau} mean-field theory
arises from the fractionalization of the composite order
parameter due to the disentanglement between the spin and orbital
degree of freedom at critical point, which was not captured
by the mean-field theory.
More interestingly, the fractionalization of the order parameter
is quiet similar to the deconfinement critical phenomena
\cite{Senthil}.

As well known, the orbital phenomena in most of spin-orbital
compounds are relevant to lattice distortion, which is generally
believed to cause the first-order phase transition in many
transition-metal oxides. The present results show that the
essential of the phase transition in spin-orbital interacting
system deviates from the second-order, and more close to the first
order. We expect more experiments to testify our prediction.
Also, we realize that the present negative value for $\delta$ can
be attributed to that we just consider the first-order and the
second-order approximation in cumulant expansion approach, thus it
is expected that including higher order approximation \cite{Sudb}
will refine and provide more accurate critical exponents.

In conclusion, we have obtained the critical exponents of the
phase transition of the Ising-like spin-orbital model, and shown
that the order-disorder phase transition behaves as a weakly
first-order; and due to fractionalization of the composite order
parameter, the spin-orbital system may belong to the same
universal class as the XY model.
\\

\acknowledgements
   This work was supported by the NSFC of China and the BaiRen Project of
the Chinese Academy of Sciences (CAS). Part of numerical calculations was
performed in CCS, HFCAS.

\newpage

{\large\bf FIGURE CAPTION} \\

\noindent {\it Fig.1} Sketched 3-site cells on the triangular
lattice, see the shaded triangles.
\\
\\

{\large\bf TABLE CAPTIONS}\\

\noindent {\it Table I} Fixed points of the renormalization
equations of Ising models with $S=1/2$\cite{Niemeijer} and $1$,
present model (SO), and XY model \cite{Rogiers}.
\\

\noindent {\it Table II} Thermal eigenvalue $\lambda_{T}$,
'thermal' exponent $\alpha_{T}$, critical temperature $K_{C}$ and
critical exponent $\alpha$ of Ising models with
$S=\frac{1}{2}$\cite{Niemeijer} and $1$, present model (SO), and
XY model \cite{Rogiers}.
\\

\noindent {\it Table III}  Magnetic eigenvalue $\lambda_{H}$,
'magnetic' exponent $\alpha_{H}$ and critical exponent $\delta$ of
Ising models with $S=1/2$\cite{Niemeijer} and $1$, present model
(SO), and XY model \cite{Rogiers}.

\newpage

\begin{figure}[htbp]\centering
\includegraphics[width=2.5cm,height=4.5cm,angle=270]{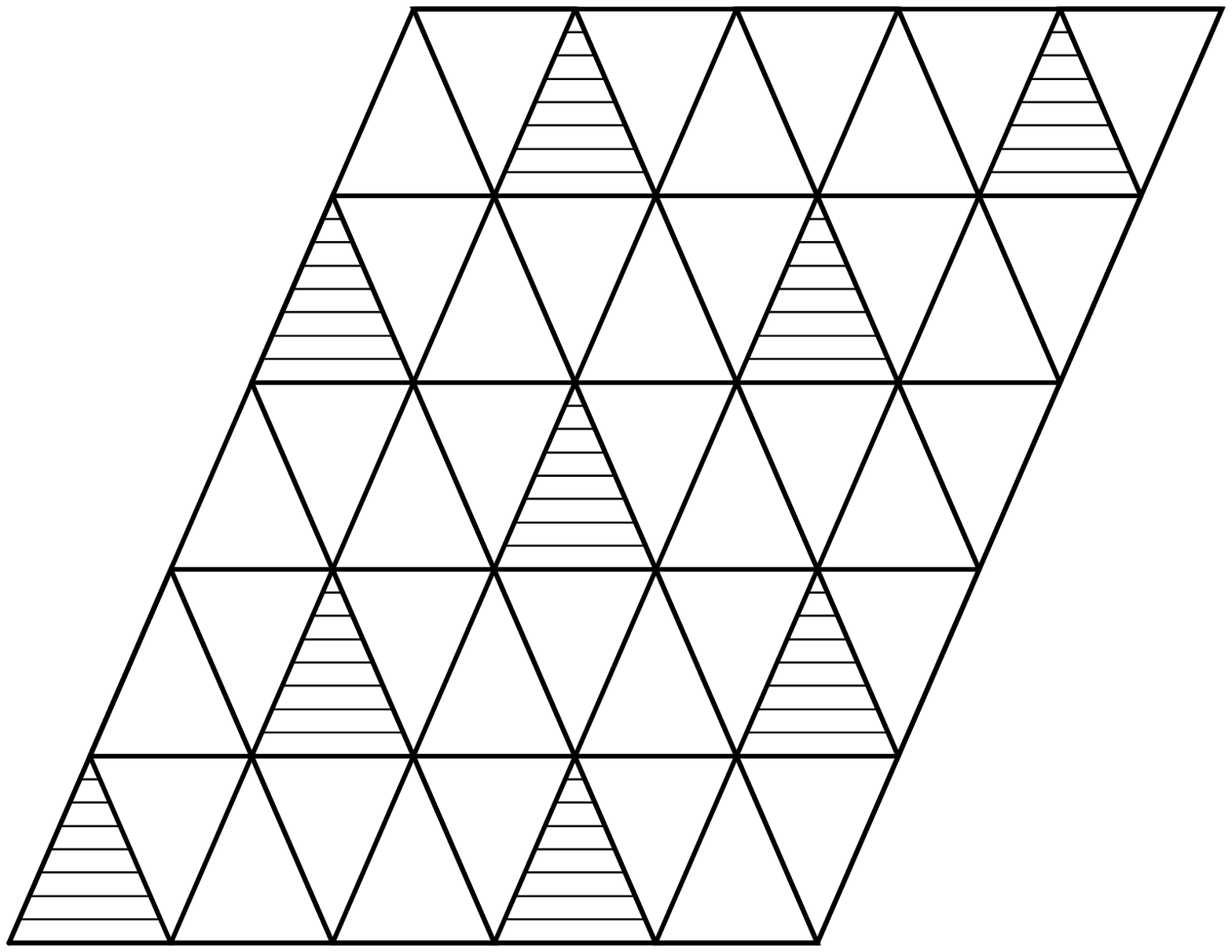}
\caption{}
\end{figure}
%
\newpage
\begin{table}
\caption{}
\begin{center}
\label{table:table}
\begin{tabular}{ll c c c c} \hline\hline
 Model         &&Ising(s=1/2)&Ising(s=1)&S-O&XY \\
 \hline
 first-order   &$K^{*}$ &0.3356&0.6312&0.4950&nonpoint\\
 \hline
 second-order  &$K^{*}$ &0.3350&0.4791&0.2568&0.8554\\
               &$L^{*}$ &-0.014&-0.011&0.0422&0.2131\\
               &$N^{*}$ &-0.015&-0.019&0.0095&-0.203\\
 \hline
\end{tabular}
\end{center}
\end{table}

\begin{table}
\caption{}
\begin{center}
\label{table:table}
\begin{tabular}{l c c c c} \hline\hline
 Model&$\lambda_{T}$&$\alpha_{T}$&$K_{C}$&$\alpha$\\
 \hline
 Ising(s=1/2)& 1.7835 & 1.8988 & 0.2514 & 0.1012 \\
 \hline
 Ising(s=1)  & 1.7909 & 1.8853 & 0.4792 & 0.1147 \\
 \hline
 S-O         & 2.4499 & 1.2260 & 0.2669 & 0.7740 \\
 \hline
 XY Model    & 2.5410 & 1.1780 & 0.8665 & 0.8015 \\
 \hline
\end{tabular}
\end{center}
\end{table}
\begin{table}
\caption{}
\begin{center}
\label{table:table}
\begin{tabular}{c c c c c} \hline\hline
 Model&Ising(s=1/2)&Ising(s=1)&S-O &XY\\
 \hline
 $\lambda_{H}$ &3.0570&3.2838&5.1410&4.9600\\
 \hline
 $\alpha_{H}$  &0.9831&0.9240&0.6710&0.6904\\
 \hline
 $\delta$      &-59.17&-13.16&-3.039&-3.230\\
 \hline
\end{tabular}
\end{center}
\end{table}

\newpage

\begin{references}

\bibitem{Kugel}
 K. I. Kugel and D. I. Khomskii, Sov. Phys. Usp {\bf 25} (1982), 231;
 Sov. Phys. JETP {\bf 37} (1973), 725.
\bibitem{Tokura}
 Y. Tokura and N. Nagaosa,  Science {\bf 288} (2000), 462.
\bibitem{Khomskii}
 D. I. Khomskii, {\it Physica Scripta} {\bf 72} (2095), CC8-14.
\bibitem{Saitoh}
 E. Saitoh, S.Okamoto, K. T. Takahashi, K. Tobe, K. Yamamoto, T. Kimura,
 S. Ishihara, S. Maekawa and Y. Tokura, Nature. {\bf 410} (2001), 181.
\bibitem{Murakami}
 Y. Murakami, H. Kawada, H. Kawata, M. Tanaka, T. Arima, Y. Moritomo, and
 Y. Tokura, Phys. Rev. Lett. {\bf 80} (1998), 1932.
\bibitem{Paul}
 P. Paul, Z.-W. Xie, R. Bau, P. D. W. Boyd, and C. A. Reed,
 J. Am. Chem. Soc. {\bf 116} (1994), 4145.
\bibitem{Choi}
 J. Choi, P. A. Dowben, S. Pebley, A. V. Bune, and S. Ducharme,
 Phys. Rev. Lett. {\bf 80} (1998), 1328;
 J. Choi, P. A. Dowben, C. N. Borca, S. Adenwalla, A. V. Bune, and S.
 Ducharme, Phys. Rev. B {\bf 59} (1999), 1819.
\bibitem{Keimer}
 B. Keimer, D. Casa, A. Ivanov, J. W. Lynn, M. V. Zimmermann,
 J. P. Hill, D. Gibbs, Y. Taguchi, and Y. Tokura,
 Phys. Rev. Lett. {\bf 85} (2000), 3946.
\bibitem{Ulrich}
 C. Ulrich, G. Khaliullin, J. Sirker, M. Reehuis, M. Ohl, S. Miyasaka,
 Y. Tokura, and B. Keimer, Phys. Rev. Lett. {\bf 91} (2003), 257202.
\bibitem{Oles}
 A. M. Oles, P. Horsch, L. F. Feiner, and G. Khaliullin,
 Phys. Rev. Lett. {\bf 96} (2006), 147205.
\bibitem{Chen}
 D.-M. Chen  W.-H. Wang and Liang-Jian Zou, {\it cond-mat/0605378}.
\bibitem{Pen}
 H. Pen, J. V. D.Brink, D. I. Khomskii, and G. A. Sawatzky
 Phys. Rev. Lett. {\bf 78} (1997), 1323.
\bibitem{Li}
 Y.-Q. Li, M. Michael, D.-N. Shi, and F. C. Zhang, Phys. Rev. Lett.
 {\bf 81} (1998), 3527.
\bibitem{Zou}
 Liang-Jian Zou, M. Fabrizio, M. Altarelli, {\it preprint}.
\bibitem{Shen}
 G.-M. Zhang and S.-Q. Shen, Phys. Rev. Lett. {\bf 87} (2001),157201.
\bibitem{Baxter}
 R. J Baxter, J. Phys. C (1973), {\bf L445}.
\bibitem{Nienhuis}
 B. Nienhuis, A. N. Berker,E. K. Riedel and M. Schick, Phys. Rev.
 Lett. {\bf 43} (1979), 737.
\bibitem{Bethe}
 H. A. Bethe,  Proc. Roy. Soc.(London)  {\bf A150} (1935), 552.
\bibitem{Niemeijer}
 T. Niemeijer and J. M. J. Van Leeuwen, Phys. Rev. Lett. {\bf
 31},1411 (1973); Physica, {\bf 71} (1974), 17.
\bibitem{Rogiers}
 J. Rogiers and R. Dekeyser, Phys. Rev. B {\bf 13} (1976), 4886.
\bibitem{Senthil}
 T. Senthil, A. Vishwanath, L. Balents, S. Sachdev, and M. P. A.
 Fisher, Science {\bf 303} (2004), 1490;
 A. J. Schofield  Science {\bf 315} (2007), 945.
\bibitem{Sudb}
 A. S. Sudb and P. C Hemmer, Phys. Rev. B {\bf 13} (1976), 980; S. Hsu,
 T. Niemeijer, and J. D. Gunton, Phys. Rev. B {\bf 11} (1975), 2699.

\end{references}
\end{document}